\documentclass[sigconf]{acmart}

\usepackage{amsmath}
\usepackage{url}
\usepackage{graphicx}
\usepackage{color}
\usepackage{enumitem}
\usepackage{booktabs}
\usepackage[normalem]{ulem}
\usepackage{wasysym}
\usepackage{microtype}
\usepackage{amsmath}
\usepackage{wasysym}
\usepackage{xspace}
\usepackage{graphicx}
\usepackage{booktabs} 
\usepackage{subcaption}
\usepackage{verbatim}
\usepackage[belowskip=1pt,aboveskip=1pt]{caption}
\newcommand{\mxnet}{MXNet\xspace} 
\newcommand{\phub}{PHub\xspace}

\newcommand{\code}[1]{\texttt{#1}}
\newcommand{\phubbox}{PBox\xspace}
\newcommand{\pbox}{PBox\xspace}

\renewcommand\footnotetextcopyrightpermission[1]{} 
\acmArticle{4}
\acmPrice{15.00}

\begin{document}
\title{Parameter Box: High Performance Parameter Servers for Efficient Distributed Deep Neural Network Training}

\author{Liang Luo$^*$,  Jacob Nelson$^\dagger$, Luis Ceze$^*$, Amar Phanishayee$^\dagger$, Arvind Krishnamurthy$^*$ \\
	$^*$University of Washington, $^\dagger$Microsoft Research}

\begin{abstract}

Most work in the deep learning systems community has focused on faster inference, but arriving at a trained model requires lengthy experiments. Accelerating training lets developers iterate faster and come up with better models.

DNN training is often seen as a compute-bound problem, best done in a single large compute node with many GPUs. As DNNs get bigger, training requires going distributed. Distributed deep neural network (DDNN) training constitutes an important workload on the cloud. Larger DNN models and faster compute engines shift the training performance bottleneck from computation to communication. Our experiments show existing DNN training frameworks do not scale in a typical cloud environment due to insufficient bandwidth and inefficient parameter server software stacks. 

We propose \phubbox, a balanced, scalable central PS hardware that balances compute and communication resources, and \phub, a high performance parameter server (PS) software design that provides an optimized network stack and a streamlined gradient processing pipeline to benefit common PS setups to utilize \phubbox. We show that in a typical cloud environment, \pbox can achieve up to 3.8x speedup over state-of-the-art designs when training ImageNet. We discuss future directions of integrating \pbox with programmable switches for in-network aggregation during training, leveraging the datacenter network topology to reduce bandwidth usage and localize data movement.

\end{abstract}

\maketitle

\section{Distributed DNN training is communication bound}
The goal of this work is to accelerate distributed DNN training in cloud environments. This work focuses on ``data'' parallelism, where workers process different samples and share the same model. A training iteration in this paradigm has two main components: computation-heavy forward and backward passes, and a communication-heavy model update step. As DNN models get larger and speedier accelerators
emerge, \textit{the performance bottleneck of distributed DNN
training has shifted from computation to communication.}

\begin{figure}[h!]
	\centering
	\begin{subfigure}[t]{0.55\columnwidth}
		\centering
		\includegraphics[width=\linewidth,trim=2 2 2 2,clip]{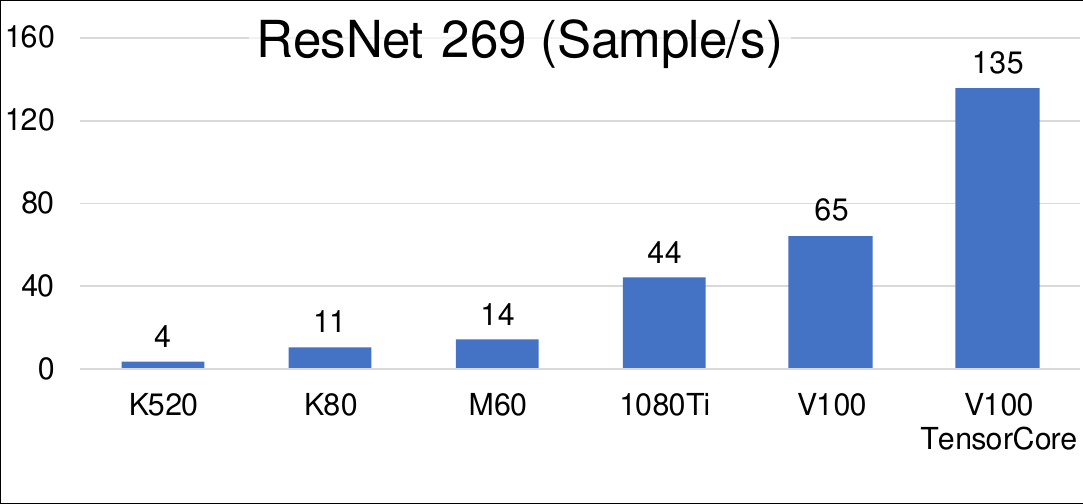}
		\caption{Per-chip GPU throughput for ResNet 269 training in EC2 has increased by 35x since 2012.}
		\label{fig:GPUPower}
	\end{subfigure}%
	\quad
	\begin{subfigure}[t]{0.4\columnwidth}
		\centering
		\includegraphics[width=\linewidth,trim=2 2 2 2,clip]{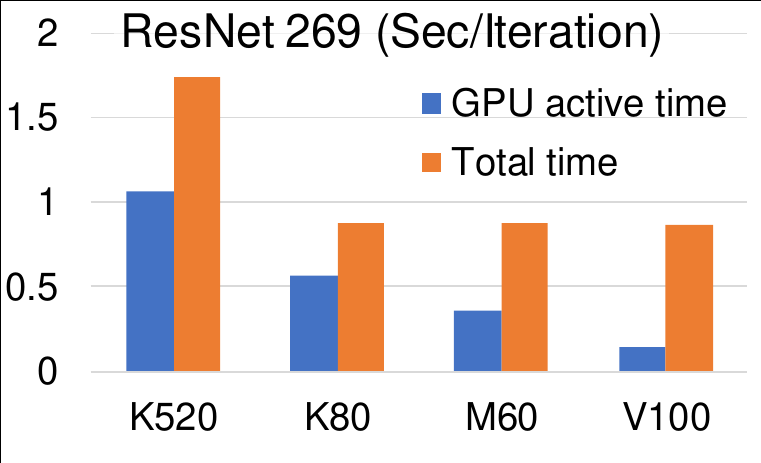}
		\caption{Increased network overhead in training as GPUs get faster.}
		\label{fig:IncreasingCommToCompRatio}
	\end{subfigure}
	\caption{Distributed training as a communication bound workload in the cloud.}
	\vspace{-1em}
\end{figure}

Larger DNN models require more gradient communication per iteration. The throughput of GPUs on ResNet, a recent DNN, has increased by 35x on modern cloud-based GPUs (Figure \ref{fig:GPUPower}), effectively demanding a similar increase in network bandwidth given a fixed batch size. However, the network bandwidth in compute instances on major cloud providers such as EC2 has not improved across generational upgrades \cite{ec2BW}. Further, existing parameter exchange mechanisms have problems scaling up the total throughput on a standard cloud network stack (Table \ref{table:frameworkPerf}). The compound effect of these factors dramatically increases communication overhead during DDNN training. 

Figure \ref{fig:IncreasingCommToCompRatio} summarizes the throughput of modest-scale DNN training with 8 workers and 8 colocated PSs on EC2 with 10Gbps links and a per GPU batch size of 4 (maximizing GPU memory usage on GRID 520): although modern DNN training frameworks can overlap backward passes with model updates, they can no longer hide the latency of communication due to faster computation. One solution is to increase the per GPU batch size, leading to a larger global batch size given a fixed number of GPUs. Large global batch sizes hurt statistical efficiency \cite{ImageNetIn1Hour, IBMAnnouncement,keskar2016large}; also, GPUs have limited memory. Techniques such as~\cite{chen2016training} could alleviate that pressure, but at a higher computational cost.

Communication overhead will likely worsen as the gap between computation and communication capability widens. New accelerators continue to reduce computation time, but networks are not getting faster at the same rate. Over the last 5 years, 100 Gbps networks have become available, but they pose high cost and have limited deployment.

These observations suggest that DDNN training has shifted from a compute-bound problem to one that also has a significant network-bound component. It is critical to perform model updates efficiently.

\section{optimized parameter servers}

\begin{table}
	\centering
	\begin{tabular}{|c|c|c|c|c|}
		\hline
		Framework   & Local & 2 workers & 4 workers & 8 workers\\
		\hline 
		TensorFlow   & 152  &  213  & 410    &  634 \\
		\hline
		Caffe2 & 195   &  266  &  343   &   513  \\
		\hline
		\mxnet & 190  &  187  &  375   &  \textbf{688}  \\
		\hline
	\end{tabular}
	\caption{Major DNN training frameworks have similar throughput for training ResNet-50 with SGD (in samples per second, using a 56Gbps IP-over-InfiniBand network and one GTX 1080 Ti GPU per worker).}
	\label{table:frameworkPerf}
\end{table}

Model updates are usually performed in a parameter server (PS), a key-value store for the current model ~\cite{ps0,ps1,ps2,Zhang:2014:DSM:2732977.2733001}. We base our work on \mxnet, a widely used, state of the art DDNN training framework that is known to be fast (Table \ref{table:frameworkPerf}, \cite{Chainer,1608.07249}) and supports native distributed training. Our profiling of \mxnet reveals two problems: (1) insufficient network bandwidth (more so with colocated PSs\footnote{The PS process and the worker training process reside in the same machine.} than non-colocated servers) and (2) an inefficient PS software stack. We found that data copy, aggregation, and optimization are the main bottlenecks in the model update process: they prevent the PS from scaling to higher throughput with high bandwidth networks.

We first propose \phub, a high performance PS design for DDNN training. We briefly summarize the main optimizations in \phub.

\noindent
\textbf{Network Stack~} Optimized InfiniBand support for lower network overhead, with one shot registration, zero copy, and minimized metadata, so all bandwidth is dedicated to gradient payload.

\noindent
\textbf{Aggregation and Optimization~} Fine grained key chunking (32KB) for maximized overlap of gradient processing and network transfer, and optimal load balancing in processor cores; locality-preserving, vectorized implementation of aggregator and optimizer.

\begin{figure}[t!]
	\centering
	\begin{subfigure}[t]{0.45\columnwidth}
		\centering
		\includegraphics[width=\columnwidth,trim=2 2 2 2,clip]{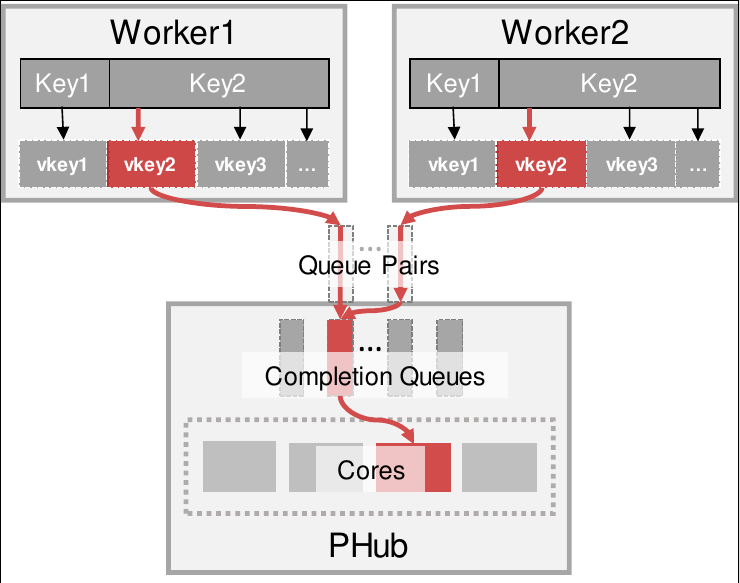}
		\caption{Fine grained key chunking and balanced chunk to core assignment scheme in \phub.}
		\label{fig:MappingChunkToCore}
	\end{subfigure}%
	\quad
	\begin{subfigure}[t]{0.45\columnwidth}
		\centering
		\includegraphics[width=\columnwidth,trim=2 2 2 2,clip]{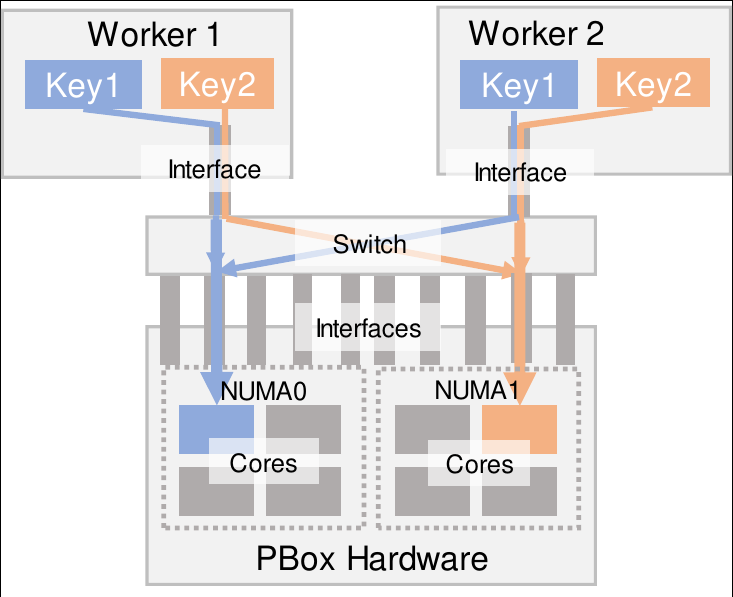}
		\caption{\phubbox has multiple NICs to balance IO, memory and network bandwidth.}
		\label{fig:PHubOverview}
	\end{subfigure}
	\caption{The \phub software and hardware architecture}
\end{figure}

\noindent
\textbf{Gradient Memory Layout~} NUMA aware, balanced scheme for assigning a key chunk to a processor core, through a series of load-balanced, locality-preserving assignment of queue pairs, interfaces, completion queues to cores (Figure \ref{fig:MappingChunkToCore}). \phub incurs zero synchronization between cores or between NUMA domains.

\begin{figure}[t!]
	\includegraphics[width=\linewidth,trim=1 1 1 2,clip]{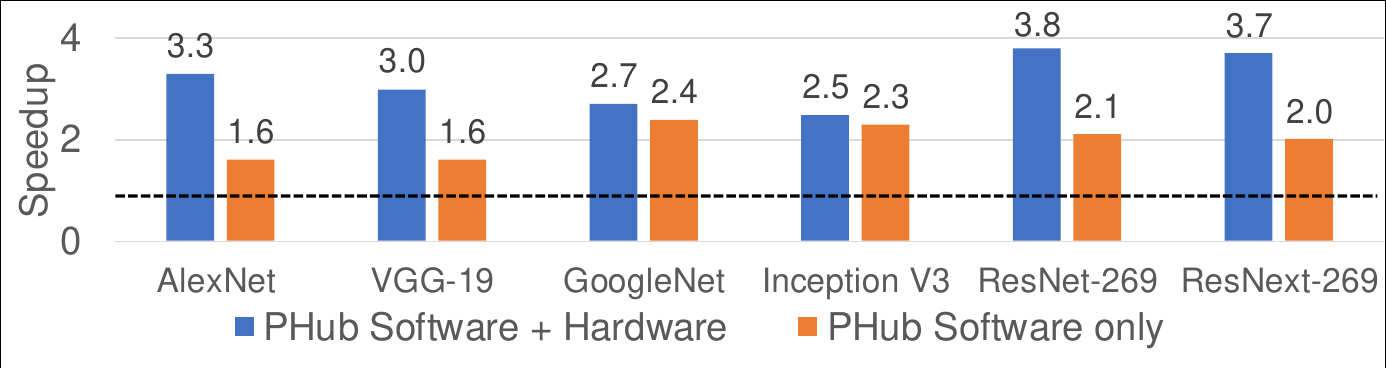}
	\caption{Training performance on a EC2-like 10Gbps network. Results are normalized to sharded \mxnet. Batch size per GPU: 8 for ResNext, 16 for ResNet, 32 for others. GPU: GTX 1080 Ti. Higher speedup possible with latest GPUs.}
	\label{fig:8GbpsRealTraining}
	\vspace{-1.5em}
\end{figure}

\begin{figure}
	\centering
	\begin{minipage}[t]{.48\columnwidth}
		\centering
		\includegraphics[width=\columnwidth, trim=3 2 2 2,clip]{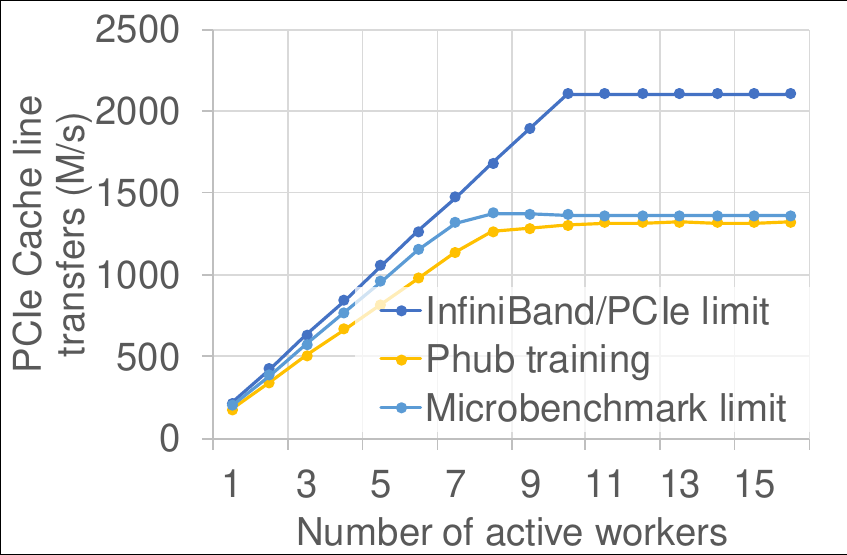}
		\caption{\phubbox can fully utilize hardware during training. Its performance is bottlenecked by the PCIe/cache controller link.}
		\label{fig:scalability}
	\end{minipage}%
	\quad
	\begin{minipage}[t]{.48\columnwidth}
		\centering
		\includegraphics[width=\columnwidth, trim=.75 1 .7 1,clip]{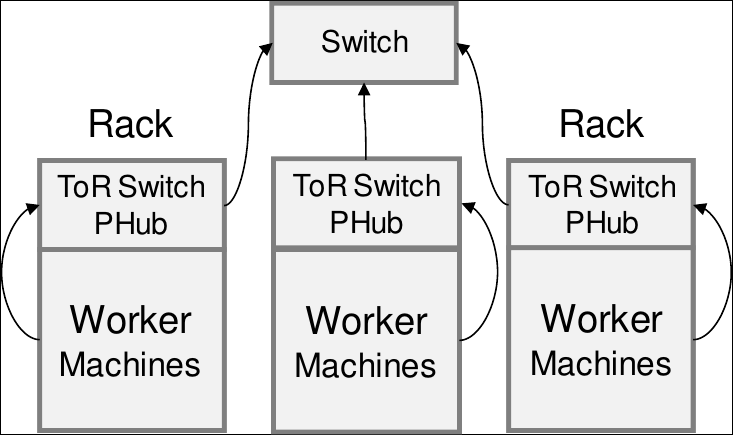}
		\caption{A potential \phub implementation with a programmable ToR switch with hybrid aggregation for workers in different racks.}
\label{fig:ina}
	\end{minipage}
\vspace{-2em}
\end{figure}


These software optimizations benefit centralized or sharded PS\footnote{Multiple PS processes, each in charge of a partition of keys.} configurations. However, to scale up a central PS, software alone is not sufficient: the hardware in a typical server is unbalanced, with significantly more computing resources than network resources. Typically, a single interface/connection in the server must handle traffic for all participating workers. We propose \phubbox, a new server architecture that balances IO, memory and network bandwidth. Our prototype \phubbox is built using a server with ten 56Gbps network interfaces---5 per NUMA domain (Figure \ref{fig:PHubOverview}). \phubbox takes full advantage of \phub software and essentially forms micro-shards inside a single box. We integrated \phub with \mxnet; Figure \ref{fig:8GbpsRealTraining} shows the speedup of \phub over a \mxnet colocated sharded baseline when training ImageNet-winning networks with 8 workers. \phub on \phubbox can provide up to 3.8x speedup over the state-of-the-art on a cloud-like 10Gbps network\footnote{All interfaces have a negotiated speed of 10Gbps with the switch in this experiment.}. The speedup using 56Gbps links is similar, ranging from 2x-7x depending on the DNN being trained.

\pbox shows linear scaling with our 8 worker cluster running all workloads, and provides higher throughput than other parameter exchange patterns using MPI or collectives (because \phub uses only one round of communication and minimum total data transfer per iteration). To understand its limits, we used a special \code{ZeroComputeEngine} that simulates infinitely fast computation in \mxnet, performing only parameter exchange operations. We found \phubbox performance is limited only by the bandwidth between the PCIe controller and the memory system (Figure \ref{fig:scalability}). This limit is hard to hit in real training: we estimate that a single \phubbox will support up to 120 workers training ResNet-50 with batch size of 32 per GPU. If each worker includes 4 GPUs, that translates to a global batch size of 15K, surpassing the maximum suggested in \cite{ImageNetIn1Hour}.
Recent work suggests larger batch sizes may impede training~\cite{ImageNetIn1Hour,IBMAnnouncement,keskar2016large,lecun1524efficient}, but if higher scalability is desired, sharding or use of new platforms with more PCIe throughput (e.g., \cite{AMDEpyc}) would enable \pbox to provide higher throughput.

\section{In-network Aggregation}


The \phubbox results show the benefit of a high-bandwidth centralized PS. Recent programmable switches~\cite{incbricks, switchkv, nopaxos} enable a new approach to building centralized PS designs to offload gradient aggregation operations to the network. Figure \ref{fig:ina} shows how \pbox architecture running in a top of rack (ToR) switch can benefit from the full bisection bandwidth inside a server rack, performing centralized aggregation of gradients inside a rack; only a single aggregated stream must be sent to higher level switches for further aggregation across racks. This hybrid synchronization reduces bandwidth usage and localizes data movement in the data center.

Current switches have limited computational capabilities: most can perform only integer operations, with little on-switch storage, and only on a small region of each packet. Our future work includes exploring the hardware requirements necessary for efficient DDNN training, as our emulation-based experiments show that these limits lead to unsatisfactory throughput on current switches.

\bibliographystyle{ACM-Reference-Format}
\bibliography{sysml} 

\end{document}